\documentclass{iopjournal}
\usepackage[T1]{fontenc}
\usepackage{lmodern,amsmath,amssymb,booktabs,microtype}
\usepackage[authoryear,round]{natbib}
\usepackage{xurl}
\newcommand{\dd}{\,\mathrm d}
\newcommand{\mrd}{\mathrm d}
\newcommand{\mri}{\mathrm i}
\newcommand{\mre}{\mathrm e}

\newcommand{\sech}{\operatorname{sech}}
\newcommand{\csch}{\operatorname{csch}}
\newcommand{\artanh}{\operatorname{arctanh}}
\newcommand{\sgn}{\operatorname{sgn}}

	\usepackage[T1]{fontenc}
	\usepackage{lmodern,amsmath,amssymb,booktabs}

\numberwithin{equation}{section}
\hypersetup{pdftitle={At last, an explicit singular value decomposition for Electrical Impedance Tomography},pdfauthor={Lionheart, Pidcock and Sebu}}
\begin{document}
\noindent\textsf{SUBMITTED}\par\bigskip
\title{At last, an explicit singular value decomposition for electrical impedance tomography}

\author{William R B Lionheart$^{1,*}$, Michael Pidcock$^2$ and Cristiana Sebu$^3$}

\affil{$^1$Department of Mathematics, The University of Manchester, United Kingdom}

\affil{$^2$School of Technology, Design and Environment, Oxford Brookes University, United Kingdom}

\affil{$^3$Department of Mathematics, University of Malta, Malta}

\affil{$^*$Corresponding author.}

\email{bill.lionheart@manchester.ac.uk}


\begin{abstract}
	\textbf{Objective.} We derive a diagonalization of the normal operator of the linearized electrical impedance tomography (EIT) problem on the disk in the case of a radially symmetric Schr\"odinger potential with Dirichlet-to-Neumann data. 
	
	\textbf{Approach.} Numerical Singular Value Decompositions (SVD) of the Electrical Impedance Tomography problem have been used since the 1980s to understand the exponential ill-posedness and non-uniform spatial resolution of the linearized problem. We find an explicit form of the normal operator, in the rotationally symmetric case,  and transform this to the adjacent-interval finite Hilbert transform, which has a known continuous spectrum and generalized eigenfunctions.
	
	\textbf{Main results.}
	The linearized operator is not compact and the spectrum is continuous, but in this spectral parameter we see the exponential decay expected from previous numerical results. Indeed we get an explicit form of the diagonal operator $\pi \sech \pi \mu$ where $\mu$ is the spectral parameter.  The generalized singular functions take the form of  complex waves with a change of variables that explains the accumulation of oscillations at the boundary.  When  the Schr\"odinger potential known in an annular neighbourhood of the boundary compactness is restored. In this case the singular values are discrete but we do not have an explicit expression for them.	
	
	\textbf{Significance.} The analysis presented in this paper is a special case: the two dimensional disk, rotational symmetry, Schr\"odinger potential rather than conductivity, Dirichlet-to-Neumann data. However it is the first explicit diagonalization of a linearized EIT problem and it illustrates the exponential ill-posedness of the problem  as well as the decay in resolution as one moves away from the boundary --- both long observed in numerical studies. 
\end{abstract}
\section{Introduction}

The singular value decomposition (SVD) of the matrix of the Fr\'echet  derivative of the forward problem of EIT has played an important role in studying the ill-posedness and non-uniformity of resolution  since the 1980s.  \cite{murai1985electrical} used a numerical SVD for their resistor mesh/FEM model. Two of the authors, WL and MP, also calculated the SVD using a FEM model and in \cite{breckon1988some} used the singular values to show the relative unimportance of drive electrode pair separation.   \cite{connolly1988inverse} published a plot of the singular values for the rotationally symmetric case. WL showed contour plots of the singular functions at the Third European Workshop on EIT in Copenhagen in 1990 and in his thesis \cite{breckonthesis}.  At  EIT2021  WL and Adler,  \cite{LionheartAdler2021}, drew attention to some anomalous results in numerical calculations of singular functions attributed to the lack of compactness, as well as commenting that the appearance of the singular functions was very similar to Zernike disk functions but compressed towards the boundary \cite{lionheart2021svd}. For an overview of the use of the SVD in  inverse problems, especially in the related X-ray tomography problem,  see \cite{hansen2021computed}.

A separate strand of work focused on finding a basis in which the matrix of the derivative operator is upper triangular rather than diagonal in a Zernike basis: \cite{allers1991stability}  and more recently  \cite{garde2024linearized} and \cite{autio2025linearization}.

In related imaging problems, such as X-ray tomography, the singular value decomposition of the Radon transform on the disk  is known for the continuum operator \cite[Ch 7 and 10]{hansen2021computed}.  Without discretizing the domain, the singular values form a discrete spectrum accumulating only at zero.  Well-designed discretizations have singular values converging to these. In the Radon transform case the discrete spectrum is a consequence of the operator being compact, which is common in many inverse problems.  In this paper we  show that in the linearized EIT case the operator is not compact, and while there is a spectral resolution analogous to the SVD  and the singular values are continuous. Nevertheless,   a family of orthogonal functions can be found, depending on this continuous parameter with an exponential decay. In the special case under consideration  we have an explicit formula for the singular functions in terms of hypergeometric functions.  To some extent the progress has been enabled by twenty-first-century developments in spectral theory we could not have anticipated in the 1980s, specifically the application of the  theory of the  Riemann-Hilbert problem to the finite Hilbert transform. However in retrospect classical spectral theory is sufficient for a basic understanding of the problem.

Here we consider the rotationally symmetric, linearized Schr\"odinger potential formulation of EIT on the unit disk. A change of variables connects its normal operator to a finite Hilbert transform and this gives a direct route to the spectral results of \citet{KatsevichTovbis2016}. The connection is also relevant to the medical-imaging audience: finite Hilbert transforms arise in interior computed tomography (CT)  after differentiated backprojection, and the relative positions of their input and output intervals determine the stability of recovery \citep{KatsevichTovbis2012}.

The distinguishing feature is a continuous spectral parameter. The generalized singular functions are not square-integrable eigenvectors, although finite numerical models necessarily have discrete singular systems. Their radial oscillations accumulate near the physical boundary. We explain both facts through a stretched radial coordinate and a scattering equation. We then consider the change when the unknown is confined to the interior.

The complete radial and angular
diagonalization is presented 
in the companion preprint \citep{Lionheart2026Disk}. The present paper focusses on the rotationally symmetric problem as presented at the EIT2026, since this produces the same spectral decay \cite{Lionheart2026AtLast}.  The full case is more complicated but retains the same underlying structure.

For a detailed survey of EIT reconstruction in the medical imaging context see \cite{LionheartAdler2021}. For a more mathematical survey and other applications see \cite{adler2015electrical}.

\section{The radial model and its normalization}
Let $\mathbb{D} = \{ (r,\theta)| 0\le r \le 1\}$ be the unit disk  where  $(r,\theta)$ are polar coordinates.
For a smooth positive conductivity $\gamma=\gamma(r,\theta)$, and electric potential $U$ we have the conductivity equation
\begin{equation}\label{eq:cond}
 \nabla\cdot(\gamma\nabla U)=0
 \end{equation}
and usually we consider the boundary data to be pairs $(U|_{\partial \mathbb{D}}, \gamma \partial U/ \partial r|_{\partial \mathbb{D}})$, of voltage and current at the boundary, from which we attempt to find $\gamma$ in $\mathbb{D}$.  The
 substitution $u=\sqrt\gamma\,U$ transforms  (\ref{eq:cond}) to
\begin{equation}
	\Delta u+qu=0,\mbox{ where } q=-\frac{\Delta\sqrt\gamma}{\sqrt\gamma}
\end{equation}
following the mathematical convention $\Delta=\nabla^2$.  We define the Dirichlet-to-Neumann map
\begin{equation}
	\Lambda_q  (u|_{r=1}) = \left.\frac{\partial u}{\partial r}\right|_{r=1}.
\end{equation}
Commonly $q$ is referred to as a Schr\"{o}dinger potential, although this terminology is used simply because the same equation arises as the stationary case in quantum mechanics, and the change of variables is sometimes called a  Liouville transformation.  Its earliest known appearance in the EIT literature appears to be \cite{kohn1984identification} and  $q$ can be thought of as an ``edge enhanced'' version of the conductivity but it has no known interpretation in electrical terms. For the purposes of qualitative imaging one could display an image of $q$, or translate to conductivity or resistivity if that was more useful.

Following  \cite{Calderon1980}
we linearize the potential-to-Dirichlet-to-Neumann map at $q=0$  and denote the potential perturbation by $\eta$. With boundary modes $e_k(\theta)=e^{ik\theta}/\sqrt{2\pi}$, the background $q=0$ solutions are $u_k(r, \theta)=r^{|k|}e_k(\theta)$. The Fr\'echet derivative as an infinite  matrix has entries
\begin{equation}
	\langle D\Lambda_0[\eta]e_k,e_l\rangle
	=-\int_{\mathbb D}\eta u_k\overline{u_l}\, r \mrd r \mrd \theta, 
	\label{eq:derivative}
\end{equation}
where $D\Lambda_0[\eta]$ denotes the Fr\'echet (total) derivative of $q \mapsto \Lambda_q$ at zero evaluated on the perturbation $\eta$. 

For radial $\eta=\eta(r)$, only diagonal entries remain and the entries for $k=n$ and $k=-n$ are the same. First we change the variable to  $t=r^2$ and $f(t)=\eta(\sqrt t)$. 
Then, we define the moment operator
\begin{equation}
A:L^2(0,1)\longrightarrow\ell^2(\mathbb N_0),
\qquad
(Af)_n=\int_0^1 f(s)s^n\,\mrd s,\quad n=0,1,\ldots.
\end{equation}
Thus \((Af)_n=-2\langle D\Lambda_0[\eta]e_n,e_n\rangle\).  One copy of each independent moment is retained, since the boundary modes \(n\) and \(-n\) give the same measurement. Equivalently, the redundant sequence \(d_k=(Af)_{|k|}\), $k\in \mathbb{Z}$,  has the multiplicity-balanced norm
\[
\|d\|_{\mathrm{mb}}^2
=|d_0|^2+\frac12\sum_{k\ne0}|d_k|^2
=\sum_{n=0}^{\infty}|(Af)_n|^2.
\]
The relative weight of the entry corresponding to $k=0$  is therefore twice that of the other entries.

With the usual \(L^2\) and one-sided \(\ell^2\) inner products, the adjoint is
\[
(A^*c)(t)=\sum_{n=0}^{\infty}c_n t^n.
\]
This formula follows immediately for finitely supported sequences; the infinite series is understood as an \(L^2(0,1)\) limit.

Using the geometric-series identity
\[
\sum_{n=0}^{\infty}(st)^n=\frac1{1-st},
\qquad 0<s,t<1,
\]
we obtain the normal operator
\begin{equation}
\begin{aligned}
	(Kf)(t):=(A^*Af)(t)
	&=\sum_{n=0}^{\infty}t^n\int_0^1f(s)s^n\,ds\\
	&=\int_0^1\frac{f(s)}{1-st}\,ds.
\end{aligned}\label{eq:kernel}
\end{equation}
The interchange of summation and integration can  be justified for \(f\) supported away from \(s=1\).  Note that if we used all integer modes but not the balanced norm we would not get such a simple expression for the kernel function of the normal operator. This foreshadows a similar argument for the full angular and radial case in \cite{Lionheart2026Disk} where a more complicated multiplicity counting is needed.

While this  kernel function has a simple form it is not a convolution operator and the kernel function becomes unbounded as $s$ and $t$ approach 1. The operator could be more closely investigated to show that it extends to  a bounded self adjoint operator on $L^2( (0,1),\mrd t)$ functions that  is {\em not} compact.  One could then go on to carefully derive the  decomposition.  However as shown below, more changes of variables reduce our problem to an adjacent interval finite Hilbert transform, which is now well understood; in order to keep the mathematical content of this paper  easily verified we follow that approach in Sec \ref{sec:FHT}. We begin with some preliminary numerical experiments.

\section{Preliminary numerical experiments}\label{sec:prelimn}

Since the numerical SVD analysis of finite-dimensional approximations is common in EIT, we first examine the results obtained by a na\"{i}ve discretization of (\ref{eq:kernel}) 
 before turning to the continuum operator analytically. First we use Gauss-Legendre discretization with 500 quadrature points to form a symmetric  Nystr\"om matrix approximation of the operator $K$ in (\ref{eq:kernel}).  The eigenvalues were calculated with Matlab's {\tt eigs} function and sorted.

\begin{figure}
	\centering
	\includegraphics[width=0.7\linewidth]{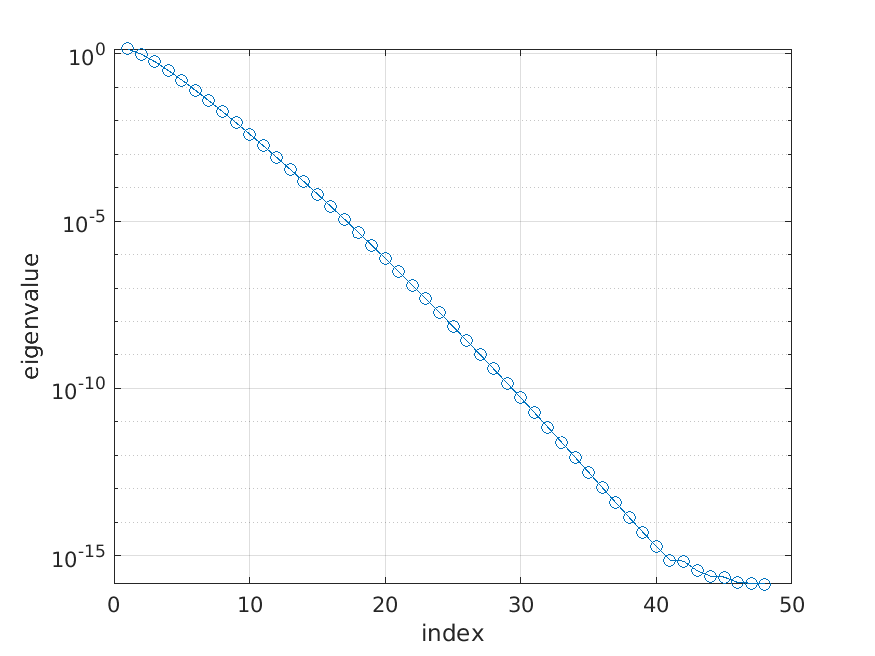}
	\caption{ Eigenvalues of the Nystr\"om matrix for the operator $K$ with 500 quadrature points. }
	\label{fig:singvalskst}
\end{figure}

\begin{figure}
	\centering
	\includegraphics[width=0.7\linewidth]{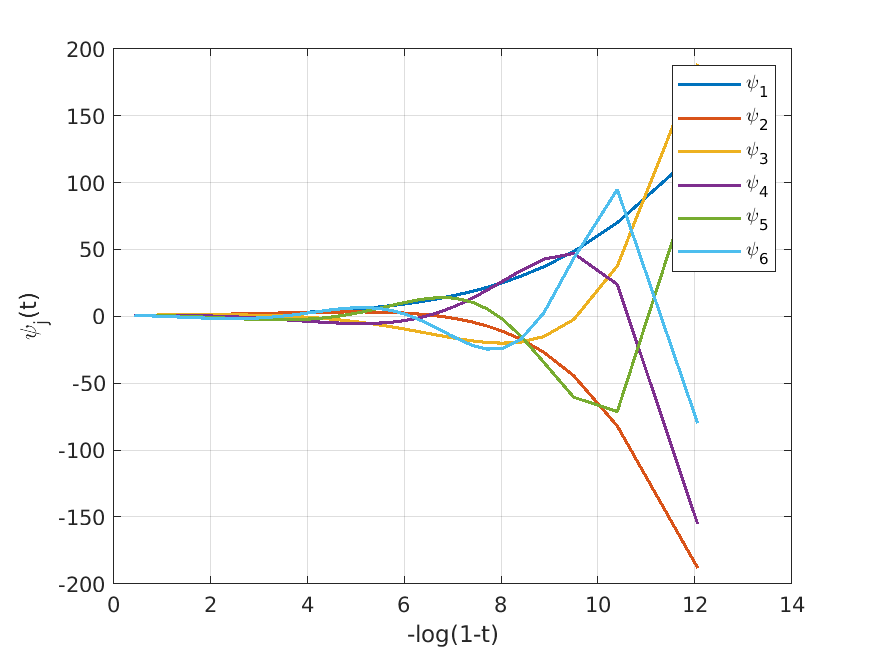}
	\caption{First six numerically calculated eigenvectors of the Nystr\"om matrix plotted on a -log(1-t) scale to show oscillations near boundary. However the quadrature rule used equally spaced samples in $t$ so the oscillations are poorly resolved}
	\label{fig:sfunslogt}
\end{figure}

\begin{figure}
	\centering
	\includegraphics[width=0.7\linewidth]{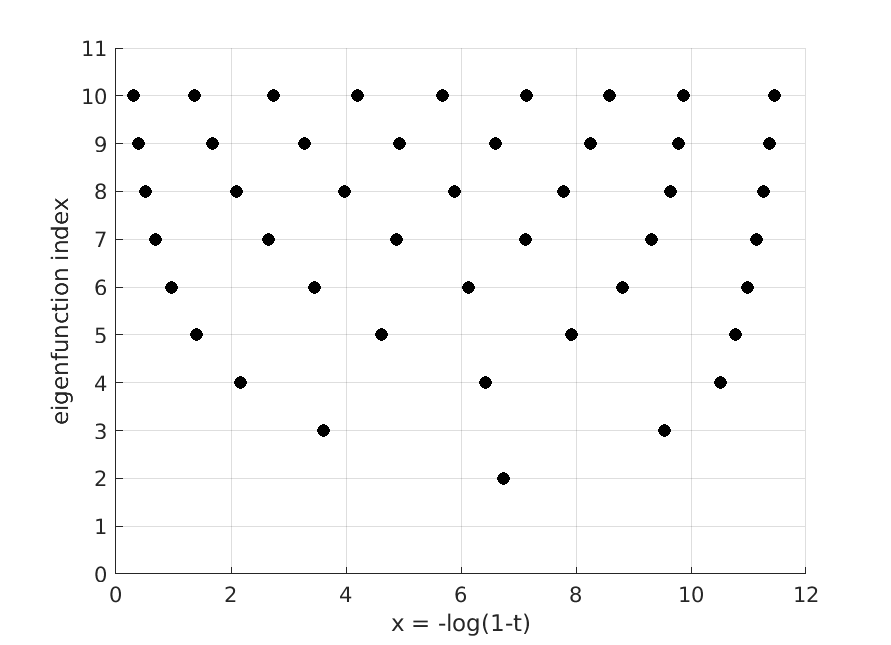}
	\caption{The zeros of the approximate eigenfunctions of the numerical eigenvectors found by interpolation and plotted on a stretched scale are approximately equally spaced.}
	\label{fig:zeros-stretched}
\end{figure}

In Fig \ref{fig:singvalskst} we see that the eigenvalues of the normal operator, and hence squared singular values of the Jacobian matrix, decay exponentiallya.   The oscillations  in the approximate right singular functions given by the eigenvectors of the matrix Fig \ref{fig:sfunslogt} show oscillations bunched towards the boundary $r=1$.  They are displayed on a stretched scale $ - \log (1-t)$ to show this more clearly. However in retrospect we should not have used equally spaced quadrature points in $t$ and the lack of smoothness at the right hand edge of the scale reflects this undersampling. In Fig \ref{fig:zeros-stretched}, the zeros of the eigenfunctions were plotted on the same stretched scale showing that they are approximately uniformly spaced. 

From these preliminary numerics we expect to see oscillations with a variable phase function given approximately by $-\log(1-t)$ together with an amplitude envelope that increases as $t$ approaches 1, and from the form of the kernel we might expect a singularity at $t=1$.  This numerical evidence might allow us to make an informed ansatz for the analytical form of the generalized singular functions and singular values but there is an added difficulty in problems with a continuous spectral parameter.  To compare  these with the numerical ones obtained here, we will have to fit the spectral parameter to the discrete ones before comparing analytical and numerical forms of the singular functions.

\section{Reduction to the finite Hilbert transform}\label{sec:FHT}
The Hilbert transform \cite[1.14(v)]{NIST:DLMF} is essentially a convolution with $1/(\pi x)$ evaluated  on the whole real line, but with the Cauchy Principal Value taken at the singularity.  It is well known in electrical engineering for its connection with causal filters and frequency modulation, and in X-ray CT from the Radon inversion formula. Its Fourier transform is $ -\mri \sgn (\omega)$ (up to choice of constants) and this is clearly straightforward to invert.  The difficulty arises with incomplete data.  We want to know the function on some interval $I$ in the real line from the data on an interval $J$. This is the finite Hilbert transform studied in \cite{KatsevichTovbis2012}, \cite{KatsevichTovbis2016}.  The key insight is that for intervals that meet at just one point, the adjacent interval case, the operator is not compact and its spectrum is a continuous interval. On the other hand for disjoint intervals with a finite gap the operator {\em is} compact and the spectrum is a discrete set of eigenvalues accumulating at zero. One of the important tools in this technique is to construct a differential operator that {\em commutes} with the Hilbert transform.  Just as commuting (diagonalizable) matrices share the same eigenspaces, finding a commuting differential operator allows us to use known eigenfunctions, or in the case of non-compact operators a family of orthogonal functions depending on a continuous spectral parameter.   That said there seems to be  no foolproof method for finding a commuting differential operator for a given integral operator, even if the kernel is given as a relatively simple algebraic formula or series. 

Our task now is to relate the operator $K$ in (\ref{eq:kernel}) to a finite Hilbert transform. Recall that the variable $t \in (0,1)$ and  we could work with the variable $1/s \in (1,\infty)$ giving adjacent intervals but one of them is infinite.  Our goal however is to transform our problem to the standard symmetric adjacent finite Hilbert transform from  $(-1,0)$ to $(0,1)$ as this has an especially nice spectral decomposition. First let   
\begin{equation}
	x=\frac{1-t}{1+t},\qquad
	(Tf)(x)=\frac{\sqrt2}{1+x}f\!\left(\frac{1-x}{1+x}\right).
	\label{eq:mobius}
\end{equation}
The amplitude factor makes $T$ unitary between the two $L^2(0,1)$ spaces. The transformed operator is
\begin{equation}
	(TKT^{-1}g)(y)=\int_0^1\frac{g(x)}{x+y}\dd x.
	\label{eq:carleman}
\end{equation}
Reflection of the input variable,  $x\mapsto -x$,   changes this to the desired finite Hilbert transform from $(-1,0)$ to $(0,1)$, up to the sign and factor $\pi$ in the conventional definition. Appendix~\ref{app:transformations} records these constants. 

The symmetric adjacent-interval result of \citet{KatsevichTovbis2016} therefore provides the continuous diagonalization. In a unitary spectral representation the normal operator acts by multiplication by
\begin{equation}
	\beta(\mu)=\pi\sech(\pi\mu),\qquad \mu\geq0.
	\label{eq:multiplier}
\end{equation}
Consequently $K$ is bounded, positive and non-compact, with purely absolutely continuous spectrum $[0,\pi]$. The forward map has generalized singular multiplier
\begin{equation}
	s(\mu)=\sqrt{\beta(\mu)}
	\sim\sqrt{2\pi}\,\mre^{-\pi\mu/2}.
	\label{eq:singular}
\end{equation}
These are multiplier functions on a continuous spectral space, not a sequence of eigenvalues. In particular, the continuous spectrum shows that the operator is not compact despite the decay of \eqref{eq:multiplier} 

A familiar example is the one-dimensional Laplacian on the whole real line. Fourier transformation turns $-\dd^2/\dd x^2$ into multiplication by $k^2$. The waves $e^{ikx}$ are not in $L^2(\mathbb R)$, but superpositions of them represent square-integrable functions. Here $\mu$ plays an analogous role. Moreover, a decaying multiplier such as $e^{-c|k|}$ is still non-compact on this continuous frequency space. The contrast is with a bounded interval and fixed endpoint conditions, where only a discrete set of frequencies is allowed.

Starting with the hypergeometric generalized eigenvalues given by \cite{KatsevichTovbis2016} and
following the transformations of the generalized eigenvalues  in Appendix A, normalized at $t=0$, we can write the generalized eigenfunctions as 
\begin{align}
	\phi_\mu(t)
	&=\frac{1}{1-t}P_{-1/2+\mri\mu}\!\left(\frac{1+t}{1-t}\right)\notag\\
	&=(1-t)^{-1/2+\mri\mu}
	{}_2F_1\!\left(\tfrac12+\mri\mu,\tfrac12+\mri \mu;1;t\right),
	\label{eq:mode}
\end{align}
where $P_\nu$ is a Legendre function and $\phi_\mu(0)=1$, and ${}_2 F_1$ is the Gauss Hypergeometric function \cite[Ch 2]{andrews1999hypergeometric}  defined by its series
\[
{}_2F_1(a,b;c;z) = \sum_{n=0}^\infty \frac{(a)_n (b)_n}{(c)_n} \frac{z^n}{n!} = 1 + \frac{ab}{c}\frac{z}{1!} + \frac{a(a+1)b(b+1)}{c(c+1)}\frac{z^2}{2!} + \cdots.
\]
 Therefore  $K\phi_\mu=\beta(\mu)\phi_\mu$ is satisfied in the generalized sense. A unitary transform also specifies a spectral measure; normalization at the origin alone is quite different from a unit norm. Although the hypergeometric expression appears at first to be complex, its value is real for $0<t<1$ and real $\mu$. See Appendix \ref{app:real}.  While the formula (\ref{eq:mode}) is explicit it is quite difficult to intuitively  understand.  Plots confirm that the solutions become highly oscillatory as $t$ approaches $1$, but in Sec \ref{sec:bdryosc} we can gain more insight from further changes of variable.

\section{Boundary oscillations and a scattering equation}\label{sec:bdryosc}
First we can change the complex power $(1-t)^{-1/2\pm \mri\mu}$ to exponential form.
Near $t=1$, the two oscillatory branches have the form: $(1-t)^{-1/2\pm \mri \mu}$
\begin{equation}
	(1-t)^{-1/2+\mri \mu}
	=(1-t)^{-1/2}\exp\!\bigl(\mri \mu\log(1-t)\bigr),
\end{equation}
which we can interpret as an amplitude envelope  and a nonlinear wave.  Real linear combinations give standing oscillations with nearly uniform zero spacing in $-\log(1-t)$ as observed in Sec \ref{sec:prelimn}. These  are spatial modes describing distinguishability in the inverse problem, not time-dependent waves propagating through tissue --- our model of quasistatic electromagnetics has ruled out propagating electromagnetic modes.

The exact coordinate and amplitude transformation is given by
\begin{equation}
	\xi=2\artanh r,\quad t=\tanh^2(\xi/2),\quad
	u_\mu(\xi)=\bigl[t(1-t)^2\bigr]^{1/4}\phi_\mu(t).
	\label{eq:liouville}
\end{equation}
The factor in \eqref{eq:liouville} preserves the $L^2$ norm under the coordinate change. In this variable the eigenequation becomes
\begin{equation}
	-u_\mu''-\frac14\csch^2\xi\,u_\mu=\mu^2u_\mu,
	\qquad 0<\xi<\infty.
	\label{eq:scattering}
\end{equation}
This is studied in the quantum scattering literature as a singular hyperbolic P\"oschl--Teller equation \cite{da2024renormalization} (for $g_c=0, g_s= -1/4$ in that paper). It is thought of as a theoretical scattering potential convenient as it has closed form solutions. At the origin the regular radial solution selects $u_\mu(\xi)\sim\sqrt{\xi/2}$, excluding the independent $\sqrt\xi\log\xi$ branch. At infinity the potential decays exponentially and the positive-energy solutions have asymptotically constant wavenumber $\mu$. The continuum interpretation is consistent with the adjacent-interval theorem. Decay of a potential alone would not exclude additional bound states for an arbitrary half-line problem; the endpoint realization is part of the statement here.

\begin{figure}[tb]
	\centering
	\includegraphics[width=\linewidth]{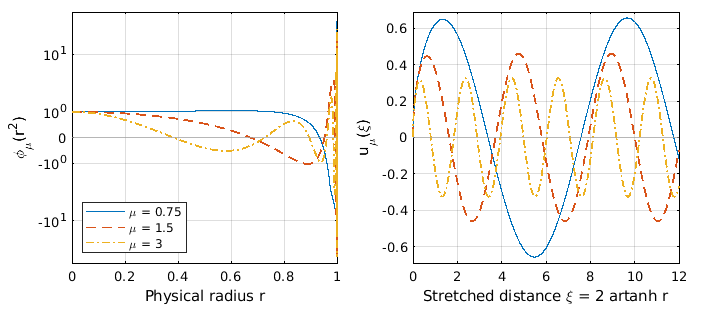}
	\caption{The same three regular generalized modes displayed in physical radius (left) and stretched distance (right). The left profiles use the normalization $\phi_\mu(0)=1$; the symmetric logarithmic vertical scale accommodates their increasing envelope. The right profiles include the norm-preserving amplitude factor in \eqref{eq:liouville}. Curves stop at $\xi=12$. These are evaluations of analytical modes calling {FCONIC} through a Matlab {\tt mex}  interface (see Sec \ref{sec:revisednumerics}) not modes of a discretized operator.}
	\label{fig:waves}
\end{figure}

For $\mu>0$, successive asymptotic zeros have separation $\pi/\mu$ in $\xi$. Since
\begin{equation}
	1-r\sim2e^{-\xi},\qquad
	-\log(1-r^2)=\xi-\log4+o(1),
\end{equation}
equal distances in $\xi$ become exponentially smaller distances in physical radius. The disk has finite Euclidean radius, but its boundary lies infinitely far away in this coordinate. This accounts for the apparent compression of oscillations without assigning a finite $L^2$ norm to an individual generalized mode.  For visualization of the solutions $\phi_\mu(r^2)$ and $u_\mu(\xi)$ for a selection of values of $\mu$ see Fig \ref{fig:waves}. 

\section{The case of the potential  known near the boundary}
Suppose the unknown perturbation vanishes for $r>R$, with $R<1$, and put $a=R^2$. The restricted moment map has normal operator
\begin{equation}
	(K_af)(t)=\int_0^a\frac{f(s)}{1-st}\dd s,\qquad 0<t<a.
	\label{eq:compact}
\end{equation}
Under \eqref{eq:mobius}, the interval $(0,a)$ becomes $(b_a,1)$, where $b_a=(1-a)/(1+a)>0$. Reflection gives the two separated intervals $(-1,-b_a)$ and $(b_a,1)$. Thus the disjoint-interval finite-Hilbert-transform theory applies \citep{KatsevichTovbis2012}. In this case $K_a$ is compact and has a discrete sequence of eigenvalues tending to zero. Their square roots are the singular values of the restricted forward map.  While the commuting differential operator and generalized singular functions are known for this problem we do not know of a closed form expression for the singular values, only asymptotic information.

The distinction between $a=1$ and $a<1$ is  geometric: an unknown extending to the boundary gives touching intervals, whereas a known boundary layer opens a gap. 

\section{Revised numerics}\label{sec:revisednumerics}

\begin{figure}
	\centering
	\includegraphics[width=0.7\linewidth]{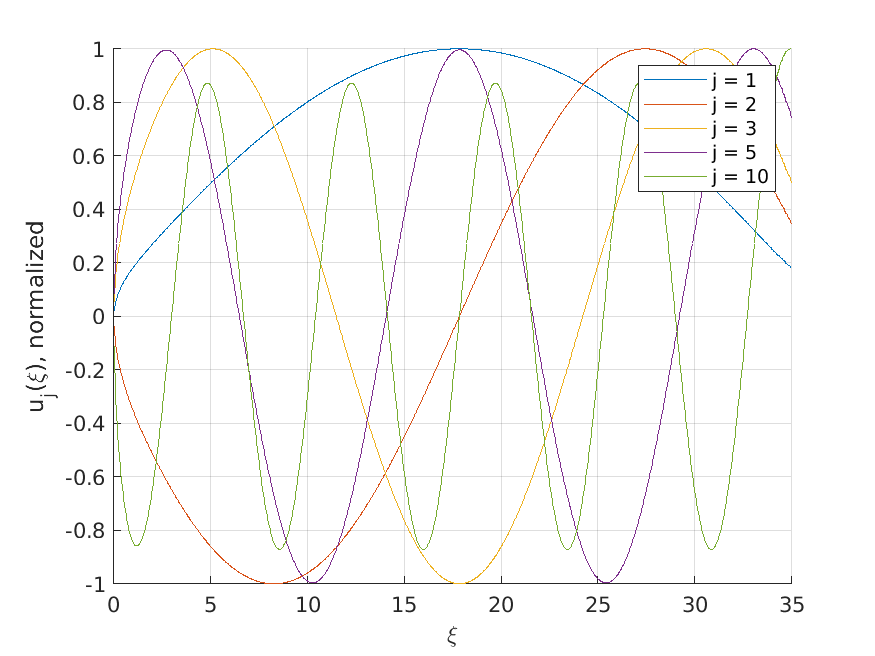}
	\caption{The eigenfunctions numbered $1, 2, 3, 5$ and $10$, of the Nystr\"om matrix for $K$ normalized to the same maximum value with quadrature points equally spaced in the $\xi$ coordinate up to a maximum value of $\xi=35$.}
	\label{fig:evalsschrodcoord}
\end{figure}

\begin{figure}
	\centering
	\includegraphics[width=0.7\linewidth]{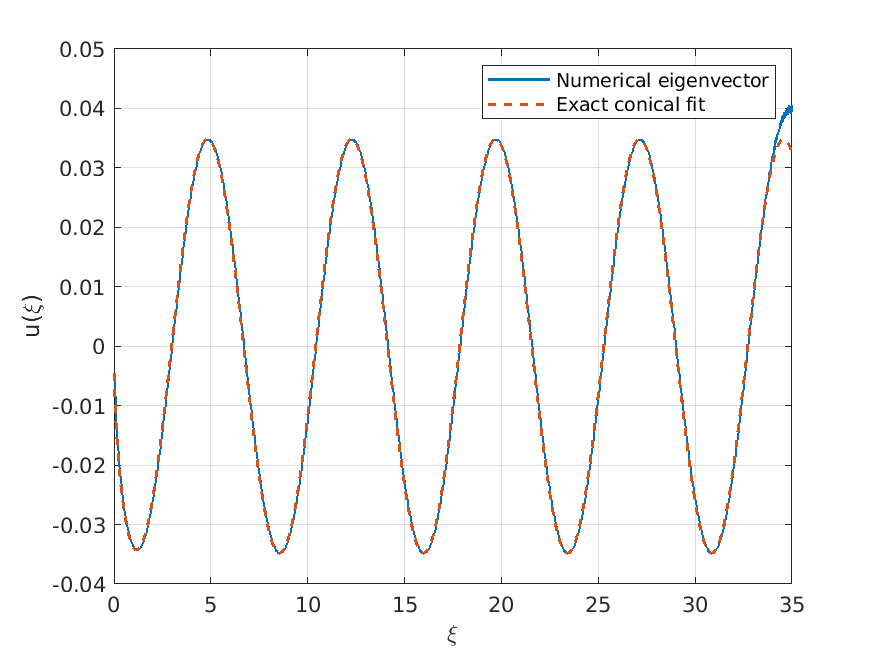}
	\caption{For the tenth numerically calculated eigenfunction using quadrature in the $\xi$ coordinate we fitted the spectral parameter as $\mu=0.84510$. Notice the noticeable discrepancy only for  large $\xi$ close to the maximum value used.}
	\label{fig:exactfit10}
\end{figure}

It is often the case that numerical experiments and theoretical developments feed each other. 
Our preliminary numerical experiments in Sec \ref{sec:prelimn}, and the earlier work in  \cite{connolly1988inverse} and \cite{breckon1988some}, provided an overall picture of what to expect. With further investigation this might have led to a more effective ansatz for the change of coordinates
needed to reveal wave-like singular functions. The process, however, has taken an unusually long time in this case—nearly 40 years. 
 Now, with the benefit of hindsight we can use a quadrature rule with points equally spaced in the $\xi$ coordinate. The results are shown in Fig \ref{fig:evalsschrodcoord}.
For comparison to our analytic expressions we need to calculate the spectral parameter $\mu$ corresponding to the  indices of the numerical eigenvectors. One way to do this is simply to estimate the wave number from zero crossings, and this works reasonably well but a more accurate method is to fit the numerical eigenvector at the quadrature points to the analytical equation for $u_\mu(\xi)$ with  $\mu$ as a parameter. We use Matlab's {\tt fminbnd}  function which finds a local minimum of a single-variable, scalar-valued function within a specified bounded interval, and is based on \cite{forsythe1977computer}. This needed many evaluations of the hypergeometric functions which were implemented  symbolically in Matlab and were consequently too slow (we were using MATLAB R2024a Update 6).  We used instead the Fortran program {\tt FCONIC} \cite{kolbig1981program}, which is specifically written for the ``conical functions'', ie $P_{-1/2 + \mri \mu}$. We  wrapped in as a {\tt mex} function to make it callable by Matlab and verified that the results agreed with Matlab's symbolic calculation of ${}_2F_1$ hypergeometric functions. See Fig \ref{fig:exactfit10} for the comparison between the 10th calculated eigenfunction and the exact formula for fitted $\mu$.

\section{Full angular and radial dependence}
A connection between the linearized EIT problem and hyperbolic geometry was anticipated even in the early work of \cite{barber1983imaging}, with  {\em backprojection along equipotential lines}  used as an approximate inverse. A connection with the generalized Radon transform in hyperbolic geometry was given by \cite{BerensteinCasadio1994}. In our present work the   coordinate $\xi=2\artanh r$ is radial distance for the Poincar\'e disk metric
\begin{equation}
	\dd s_H^2=\frac{4(\dd r^2+r^2\dd\theta^2)}{(1-r^2)^2}.
\end{equation}
In this metric the equipotential lines from a dipole source on the boundary are geodesics. 
This also gives a geometric interpretation of the half-line in the radial calculation as well as a connection to hyperbolic geometry. In the full problem with radial and angular dependence, the multiplicity-normalized normal operator is
\begin{equation}
	(K_{\mathbb D}h)(z)=\frac1\pi\int_{\mathbb D}
	\frac{h(w)}{|1-z\overline w|^2}\dd A(w).
\end{equation}
Its radial restriction is precisely \eqref{eq:kernel}. A unitary change of measure identifies it as a function of the hyperbolic Laplacian, with the same multiplier $\beta(\mu)$ \citep{Lionheart2026Disk}.

\section{Conclusions}
In this paper we diagonalized the normal operator of the rotationally symmetric linearized Schr\"odinger Calder\'on problem on the disk.  Despite this being a simplified problem it already reveals the exponential illposed nature of the problem, and rapidly decaying resolution towards the centre observed numerically.  With a change of coordinates the normal operator has a simple algebraic form and the generalized singular functions can be thought of as standing waves with a non-uniform phase accumulating at the boundary and an amplitude attenuated towards the centre. In this paper we side-step some mathematical details by transforming our problem to a well-studied finite Hilbert transform.  We note that the continuous spectrum becomes discrete as soon as we assume the conductivity (and hence Schr\"odinger potential) is known near the boundary. These features persist even when the assumption of angular invariance is removed, although mathematically more complicated. We hope that this will provide an insight into the difficulty of EIT in realistic 3D problems with more general shapes and linearization about a non-uniform background.  In cases where the rotational symmetry is realistic, our explicit diagonalization could be used to give closed-form regularized inversion formulae. 

\appendix

\section{Verification of the changes of variables}\label{app:cov}
\label{app:transformations}
Writing $t=(1-y)/(1+y)$ and $s=(1-x)/(1+x)$ gives
\begin{equation}
	\left|\frac{\dd s}{\dd x}\right|=\frac{2}{(1+x)^2},\qquad
	1-st=\frac{2(x+y)}{(1+x)(1+y)}.
\end{equation}
The two square-root Jacobian factors therefore give
\begin{equation}
	\frac{\sqrt{|\dd s/\dd x|}\sqrt{|\dd t/\dd y|}}{1-st}
	=\frac1{x+y},
\end{equation}
which proves \eqref{eq:carleman}. If
\[
(Hh)(y)=\frac1\pi\int_{-1}^0\frac{h(z)}{z-y}\dd z,
\qquad (Rg)(z)=g(-z),
\]
then $TKT^{-1}=-\pi HR$. 

For the symmetric intervals with endpoints $-1$ and $1$, the regular function in \citet[equation (4.12)]{KatsevichTovbis2016} is
\begin{equation}
	\psi_\mu(x)=x^{-1/2+\mri \mu}
	{}_2F_1\!\left(\tfrac14+\tfrac{\mri \mu}{2},\tfrac34+\tfrac{\mri \mu}{2};1;1-x^2\right).
\end{equation}
Pulling this function back with $T^{-1}$ and normalizing to one at $t=0$ gives
\begin{equation}
	\phi_\mu(t)=\frac{1+x}{2}\psi_\mu(x),\qquad x=\frac{1-t}{1+t}.
\end{equation}
The quadratic hypergeometric transformation identifies this expression with \eqref{eq:mode}. Equivalently, both expressions solve the same second-order equation below and are its regular solution with value one at $t=0$. 

For  the commuting differential operator set $p(t)=t(1-t)^2$ and
\begin{equation}
	P\phi=(p\phi')'+t\phi,\qquad D=-P+\tfrac34.
\end{equation}
The regular mode in \eqref{eq:mode} satisfies
\begin{equation}
	P\phi_\mu=(\tfrac34-\mu^2)\phi_\mu.
\end{equation}
Since $\dd t/\dd\xi=\sqrt p$, the substitution $u=p^{1/4}\phi$ is unitary and conjugates $D$ to $-\mrd^2/\mrd\xi^2-\tfrac14\csch^2\xi$ with the regular endpoint realization. The sign and the additive shift are essential.

Finally, the Legendre representation \citep[section 14.3]{DLMF} gives
\[
P_{-1/2+\mri \mu}(\cosh\xi)
={}_2F_1\!\left(\tfrac12-\mri \mu,\tfrac12+\mri \mu;1;-\sinh^2(\xi/2)\right).
\]
Applying the Pfaff transformation \cite[Thm 2.2.5]{andrews1999hypergeometric} with $t=\tanh^2(\xi/2)$ gives \eqref{eq:mode}. It also gives the stable profile
\begin{equation}
	u_\mu(\xi)=\sqrt{\frac{\sinh\xi}{2}}\,
	P_{-1/2+\mri \mu}(\cosh\xi).
\end{equation}

\section{Generalized eigenfunctions are real valued}\label{app:real}

Given 
\[
\phi_\mu(t)
=
(1-t)^{-1/2+\mri \mu}
{}_2F_1\left(
\frac12+\mri\mu,\frac12+\mri\mu;
1;t
\right),
\qquad
0<t<1,
\]
where \(\mu\in\mathbb{R}\), although both the complex power and the hypergeometric function appearing in this expression are in general complex, their phases cancel and \(\Phi_\mu\) is real-valued.

To see this, let
\[
F_\mu(t)
=
{}_2F_1\left(
\frac12+\mri\mu,\frac12+\mri \mu;
1;t
\right).
\]
Euler's transformation \cite[Thm 2.2.5]{andrews1999hypergeometric}
\[
{}_2F_1(a,b;c;t)
=
(1-t)^{c-a-b}
{}_2F_1(c-a,c-b;c;t)
\]
gives, with
\[
a=b=\frac12+\mri\mu,
\qquad c=1,
\]
gives the identity
\[
F_\mu(t)
=
(1-t)^{-2\mri \mu}
{}_2F_1\left(
\frac12-\mri\mu,\frac12-\mri \mu;
1;t
\right).
\]
For real \(t\in(0,1)\) and real \(\mu\),
\[
{}_2F_1\left(
\frac12-\mri\mu,\frac12-\mri \mu;
1;t
\right)
=
\overline{F_\mu(t)}.
\]
Consequently,
\[
F_\mu(t)
=
(1-t)^{-2\mri\mu}\overline{F_\mu(t)}.
\]
Multiplication by (\(1-t)^{\mri \mu}\) gives
\[
(1-t)^{\mri\mu}F_\mu(t)
=
(1-t)^{-\mri\mu}\overline{F_\mu(t)}
\overline{(1-t)^{\mri \mu}F_\mu(t)}.
\]
It follows that
\[
(1-t)^{\mri \mu}
{}_2F_1\left(
\frac12+\mri\mu,\frac12+\mri \mu;
1;t
\right)
\]
is real-valued. Hence
\[
\phi_\mu(t)\in\mathbb{R},
\qquad 0<t<1,\quad \mu\in\mathbb{R}.
\]
Thus the apparently complex representation
\[
\phi_\mu(t)
=(1-t)^{-1/2}
\left[
(1-t)^{\mri \mu}
{}_2F_1\left(
\frac12+\mri \mu,\frac12+\mri \mu;
1;t
\right)
\right]
\]
is in fact a representation of a real generalized eigenfunction.

\data{Source code in Matlab and Fortran is available and reproduces the figures}

\ack{ ChatGPT 4.0 (OpenAI) Plus found the connection to the P\"oschl--Teller  potential and  outlined the Matlab code used for numerical examples.  The authors take full responsibility for the correctness of the mathematics and programming. WL would like to thank the Isaac Newton Institute for Mathematical Sciences, Cambridge, for support and hospitality during the programme RNT,  and Clare Hall, Cambridge where he was a visiting fellow, when some of the work on this paper was undertaken. This work was supported by EPSRC grant EP/Z000580/1 and EP/V007742/1. WL would like to thank Alexander Katsevich and Marco Marletta for helpful discussions}
\bibliographystyle{plainnat}
\bibliography{references}
\end{document}